\documentclass{physeauth}
\usepackage{graphicx}
\usepackage{amsmath}
\usepackage{amssymb}

\begin{document}

\begin{frontmatter}

\title{
Excitation Spectrum of Bilayer $\nu=2$ Quantum Hall Systems
}

\author[thank1]{Y. Shimoda},
\author{T. Nakajima}
and
\author{A. Sawada} 

\address{Department of Physics, Tohoku University, Sendai, 980-8578, Japan}

\thanks[thank1]{
Corresponding author.\\ 
E-mail: yshimoda@lowtemp.phys.tohoku.ac.jp}

\begin{abstract}
Excitation spectra in bilayer quantum Hall systems 
at total Landau-level filling $\nu=2$ are studied
by the Hartree-Fock-Bogoliubov approximation.
The systems have the spin degrees of freedom
in addition to the layer degrees of freedom
described in terms of pseudospin.
On the excitation spectra from
spin-unpolarized and pseudospin-polarized ground state,
this approximation fully preserves the spin rotational symmetry
and thus can give not only spin-triplet but also 
spin-singlet excitations systematically.
It is also found that 
the ground-state properties are well described
by this approximation.
\end{abstract}

\end{frontmatter}

\section{Introduction}
Strong interactions often drive low-dimensional systems 
into exotic new phases. 
For a two-dimensional electron system (2DES) 
under high perpendicular magnetic fields, 
the interaction dominates the system properties 
because the kinetic energy is quenched by the Landau-level quantization.
One of the most interesting phenomena in this strongly-correlated 
system is the quantum Hall effect, which has attracted 
a great deal of experiment and theoretical interest \cite{QHE}.
Recent advances in material growth techniques have made it possible 
to fabricate high-quality 2DESs confined to two parallel layers. 
By the introduction of such layer degrees of freedom 
a lot of new correlation effects can be realized 
because the strength of interlayer interactions 
and interlayer tunneling are controllable \cite{QHE2}.

In a bilayer quantum Hall (QH) system 
at total Landau-level filling $\nu=2$,
theoretical \cite{nu2HF1}\nocite{nu2HF2}\nocite{est}--\cite{nu2ED} 
and experimental studies
\cite{nu2exp1}\nocite{nu2exp2}--\cite{nu2exp3} have confirmed 
the existence of a canted antiferromagnetic 
phase (CAF) between a fully spin-polarized ferromagnetic phase 
and a spin-singlet one.
The properties of low-lying excitations in this system 
have been discussed by the Hartree-Fock approximation (HF) \cite{nu2HF1} 
and exact diagonalization (ED) calculations \cite{nu2ED}.
However, the HF calculation neither preserves  
the spin rotational symmetry nor can well describe pseudospin correlations,
while the ED calculation is only applicable to small-size systems.
Thus, in order to investigate the excitation spectra of large-size systems,
we adopt a better approximate approach 
called the Hartree-Fock-Bogoliubov (HFB) 
approximation~\cite{ring,nuclear,HFB}, 
and then write down the effective Hamiltonian of the systems.
This approximation can take particle-hole correlations
into consideration better 
than the Hartree-Fock (HF) approximation
and further preserves the spin and pseudospin rotational symmetries 
in contrast to insufficient treatment by the HF approximation.
We discuss not only excitation spectra but also ground-sate properties
based on this approximation.
\section{Hartree-Fock-Bogoliubov Approach to Bilayer $\nu=2$ QH Systems}
In the presence of interlayer tunneling
in double-quantum-well structures,
the single-particle states are split into
symmetric and antisymmetric combinations of one-layer states. 
These layer degrees of freedom can be described 
in terms of pseudospin as $\sigma _x = \uparrow, 
\ \downarrow$.
In the case of large interlayer-tunneling energy $\Delta_{\rm SAS}$
and small layer separation $d$ (typically 
$\Delta_{\rm SAS}/E_{C} \geq 0.6$ and $d/l_{B} \leq 1.0$, 
where $E_{C}=e^2/\epsilon l_{B}$ and $l_{B}=\sqrt{c\hbar/eB}$)
considered in this paper, 
the ground state of bilayer $\nu=2$ QH system is spin-singlet and 
fully pseudospin polarized 
because the Zeeman energy is much smaller than 
the tunneling energy and the interaction energy.
For simplicity we consider only the lowest Landau levels.
We can write down the Hartree-Fock-Bogoliubov Hamiltonian 
in such case.
We consider $N$-electron systems on a sphere \cite{sphere} 
whose surface is passed through by $2S$ flux quanta,
where $2S=N/2-1$ for $\nu=2$. 
The effective Hamiltonian is given by 
\begin{eqnarray}
{\mathcal H}_{\rm eff} &=& {\mathcal H}_0 
+ {\mathcal H}_{\pm}, \label{eqn:HFBH} \\ 
{\mathcal H}_{0} &=& \sum_{K,N} \bigl\{\,(\Delta_{\rm SAS}+e_{K}+\lambda_{K})
(C_{KN}^{\dagger}C_{KN}+H_{KN}^{\dagger}H_{KN}) \nonumber \\
& &+\frac{\lambda_{K}}{2}(-1)^{N}(C_{KN}^{\dag}C_{K,-N}^{\dag}
+ H_{KN}^{\dag}H_{K,-N}^{\dag}+{\rm h.c.}) \nonumber \\
& & +f_{K}[C_{KN}^{\dagger}H_{KN}+(-1)^{N}C_{KN}^{\dagger}
H_{K,-N}^{\dagger}+{\rm h.c.}]\,\bigl\} , \label{eqn:Szero} \\ 
{\mathcal H}_{\pm} &=& \sum_{K,N}\bigl[(\Delta_{\rm SAS}
+\Delta_{\rm Z}+e_{K}+\lambda_{K}-g_{K})D_{KN}^{\dagger}D_{KN} \nonumber\\
& &+(\Delta_{\rm SAS}-\Delta_{\rm Z}+e_{K}+\lambda_{K}-g_{K})
F_{KN}^{\dagger}F_{KN} \nonumber\\
& &-g_{K}(-1)^{N}(D_{KN}^{\dagger}F_{K,-N}^{\dagger}+{\rm h.c.}) \bigl] ,
\label{eqn:Triplet} 
\end{eqnarray}

\begin{figure}[t]
\includegraphics[width=.70\linewidth]{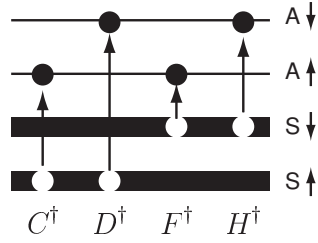}
\caption{Excitons represented by operators $C^{\dag}$, 
$D^{\dag}$, $F^{\dag}$, and $H^{\dag}$. 
Energy levels occupied by electrons 
[\,i.e., $S\uparrow$ (Symmetric and up-spin) 
and $S\downarrow$ (Symmetric and down-spin) levels\,]
are shown by thick lines, while 
solid and open circles indicate excited electrons into 
an unoccupied level [Antisymmetric 
and up-spin (or down-spin)]
and holes 
in the $S\uparrow$ (or $S\downarrow$) 
levels, respectively. 
The changes in the $z$ component of 
total spin introduced by $C$, $D$, 
$F$ and $H$ excitons are $0$, $-1$, $+1$ and $0$, respectively.
\label{fig:excitons}}
\end{figure}

\begin{equation}
e_{K} = \sum_{J=0}^{2S}(2J+1)V_{\rm intra}^{J}
\left[\frac{(-1)^{2S-J}}{2S+1}-
{S S J \brace S S K} \right], \nonumber 
\end{equation}
\begin{equation}
\lambda_{K} = \sum_{J=0}^{2S}(2J+1)
\frac{V_{\rm intra}^{J}-V_{\rm inter}^{J}}{2}
[(-1)^{2S-J}-1]{S S J \brace S S K}, \nonumber 
\end{equation}
\begin{equation}
f_{K} = \sum_{J=0}^{2S}(2J+1)
\frac{V_{\rm intra}^{J}-V_{\rm inter}^{J}}{2}
(-1)^{2S-J}{S S J \brace S S K}, \nonumber 
\end{equation}
\begin{equation}
g_{K} = f_{K}-\lambda_{K}. \nonumber
\end{equation}

Here 
${\mathcal H}_{\rm eff}$ is composed of two parts,
${\mathcal H}_0$ and ${\mathcal H}_{\pm}$,
described by such excitations 
that change the $z$ component of total spin by $0$ and $\pm 1$,
respectively.
$\Delta_{\rm SAS}$ is the energy gap between symmetric (S) 
and antisymmetric (A) single-particle states, and 
$\Delta_{\rm Z}$ is the Zeeman gap between spin up ($\uparrow$) 
and down ($\downarrow$) states.
Because the system has the spatial rotational symmetry,
the interaction matrix elements can be expressed 
in terms of Wigner's $6j$ symbol \({SSJ \brace SSK}\) and 
intra/inter-layer pseudopotentials,  
$V^{J}_{\rm intra/inter}$ for relative angular momentum $2S-J$. 
$C^{\dag}_{KN}$, $D^{\dag}_{KN}$, $F^{\dag}_{KN}$, and $H^{\dag}_{KN}$ 
are exciton creation operators (see Fig.\ref{fig:excitons}).
For example,
$C^{\dag}_{KN} = \sum_{l,m} \langle Sl;S,-m|KN \rangle 
a^{\dag}_{l\uparrow} (-1)^{S-m} s_{m \uparrow}$ and
this creates an exciton [a hole in the S $\uparrow$ level 
and a particle in the A $\uparrow$ level]
with the total angular momentum $K$ and its $z$ component $N$, 
where \(\langle Sl;S,-m|KN\rangle\) 
is the Clebsh-Gordan coefficient, and $a^{\dag}_{l\sigma}$ 
($s^{\dag}_{m\sigma^{\prime}}$) creates an electron occupying
the $l$ ($m$)-th antisymmetric (symmetric) combination of 
Landau orbits with spin-$\sigma$ ($\sigma^{\prime}$).  


The diagonalization of ${\mathcal H}_{\pm}$
in Eqn.(\ref{eqn:Triplet}) can be performed 
by the following Bogoliubov transformation:
\begin{eqnarray}
{\mathcal H}_{\pm} &=& \sum_{K,N}
\bigl[\,(\Delta_{\rm Z}+\omega_{K}^{T})
R_{KN}^{\dag} R_{KN} \nonumber \\
& & \qquad +(-\Delta_{\rm Z}+\omega_{K}^{T}) S_{KN}^{\dag}S_{KN} 
\bigl] , \label{eqn:pmH} \\
\omega_{K}^{T} &=& \sqrt{(\Delta_{\rm SAS}+e_{\rm K})
(\Delta_{\rm SAS}+e_{\rm K}-2g_{\rm K})} \,, \nonumber \\
R_{KN} &=& D_{KN}\cosh{\frac{\phi_{K}}{2}}
+(-1)^{N}F_{K,-N}^{\dag}\sinh{\frac{\phi_{K}}{2}} ,  \nonumber \\
S_{K,-N} &=& (-1)^{N}D_{KN}^{\dag}\sinh{\frac{\phi_{K}}{2}}
+F_{K,-N} \cosh{\frac{\phi_{K}}{2}} , \nonumber
\end{eqnarray}
where 
$\tanh{\phi_{K}}=-g_{K}/(\Delta_{\rm SAS}+e_{K}-g_{K})$ 
and $\omega_{K}^{T}$ gives the energy of spin-triplet excitation.

The Hamiltonian ${\mathcal H}_{0}$ 
in Eqn.(\ref{eqn:Szero}) 
can be decomposed into a spin-triplet part and a spin-singlet one. 
In fact, as linear combinations of operators $C_{KN}$ and $H_{KN}$,
a new set of operators, $Q_{KN}$ and $P_{KN}$, 
can be introduced as
$P_{KN} = (C_{KN}-H_{KN})/\sqrt{2}$, 
$Q_{KN} = (C_{KN}+H_{KN})/\sqrt{2}$,
and then ${\mathcal H}_{0}$ can be written 
in terms of $Q_{KN}$ and $P_{KN}$
in the following form:
\begin{eqnarray}
{\mathcal H}_0&=&{\mathcal H}_{\rm triplet}^{0}
+{\mathcal H}_{\rm singlet}, 
\label{eqn:Szero2} \\
{\mathcal H}_{\rm triplet}^{0}&=&\sum_{K,N}\bigl[\,
(\Delta_{\rm SAS}+e_{K}-g_{K}) P_{KN}^{\dag} P_{KN} \nonumber \\
& &-\frac{g_{K}}{2} (-1)^{N} (P_{KN}^{\dag}P_{K,-N}^{\dag} 
+ {\rm h.c.} ) \bigl] , 
\label{eqn:triplet2} \\
{\mathcal H}_{\rm singlet} &=& \sum_{K,N} \bigl [
(\Delta_{\rm SAS}+e_{K}+2\lambda_{K}
+g_{K}) Q_{KN}^{\dag} Q_{KN} \nonumber \\
& & + \frac{2\lambda_{K}+g_{K}}{2} (-1)^{N} 
(Q_{KN}^{\dag} Q_{K,-N}^{\dag}+ {\rm h.c.})
\bigl]. 
\label{eqn:CDW}
\end{eqnarray}
Each part in Eqn.(\ref{eqn:Szero2}) 
can be diagonalized by the following Bogoliubov 
transformations, respectively, as 
\begin{eqnarray}
{\mathcal H}_{\rm triplet}^{0} 
&=& \sum_{K,N} \omega_{K}^{T}\,T_{KN}^{\dag}T_{KN}, \\
{\mathcal H}_{\rm singlet} &=&
\sum_{K,N} \omega_{K}^{S} \,U_{KN}^{\dag} U_{KN}, \\
\omega_{K}^{S} &=& \sqrt{(\Delta_{\rm SAS}+e_{K})
(\Delta_{\rm SAS}+e_{K}+4\lambda_{K}+2g_{K})} \,, \nonumber \\
T_{KN} &=& P_{KN}\cosh{\frac{\phi_{K}}{2}}
+(-1)^{N}P_{K,-N}^{\dag}\sinh{\frac{\phi_{K}}{2}}, \nonumber \\
U_{KN} &=& Q_{KN}\cosh{\frac{\varphi_{K}}{2}}
+(-1)^{N}Q_{K,-N}^{\dag}\sinh{\frac{\varphi_{K}}{2}} , \nonumber 
\end{eqnarray}
where 
$\tanh{\varphi_{K}}=(2\lambda_{K}+g_{K})/
(\Delta_{\rm SAS}+e_{K}+2\lambda_{K}+g_{K})$,
$\omega_{K}^{S}$ is the energy of spin-singlet excitation.
The definitions of $\omega_{K}^{T}$ and $\phi_{K}$ 
have already been given 
on the diagonalization of ${\mathcal H}_{\pm}$ 
in Eqn.(\ref{eqn:pmH}). 

\section{Results and Discussion}

\begin{figure}[t]
\includegraphics[scale=.60]{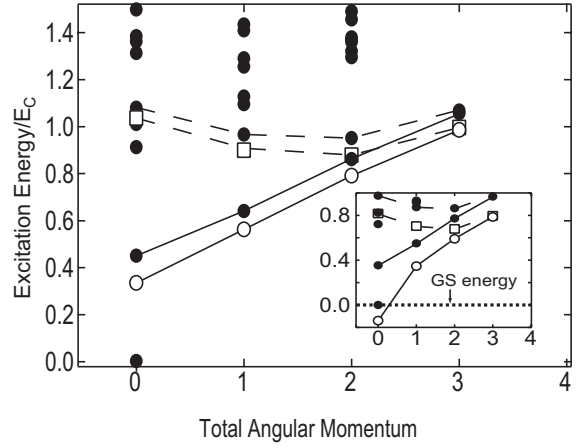}
\caption{Low-lying excitation spectrum as a function of total
angular momentum is shown for eight electrons, $d/l_{\rm B}=1$, and 
$\Delta_{\rm SAS}/E_{\rm C}=0.7$.
Open circles (squares) linked by solid (dashed) line
show the spin-triplet (-singlet) excitation spectrum 
by the HFB approximation.
Excitation spectrum obtained by the ED method
are shown by solid circles, and 
spin-triplet (-singlet) ones obtained by the ED method 
are linked by solid (dashed) lines 
as a guide to the eye.
In the inset, the low-lying excitation spectrum for 
$\Delta_{\rm SAS}/E_{\rm C}=0.6$ is shown.
\label{fig:spectrum}}
\end{figure}

In Fig.~\ref{fig:spectrum}
we show calculated results of excitation spectra
in eight electron systems with $d/l_{B}=1.0$.
Spin-triplet and spin-singlet excitation energies,
$\omega_{K}^{T}$ and $\omega_{K}^{S}$, by HFB approximation are 
shown by open circles and squares, respectively.
Calculated spectrum by the ED method is also shown by solid circles.
Spin-triplet and spin-singlet excitations 
obtained by the ED method are linked by
solid and dashed lines, respectively, as a guide to the eye.
In the figure the contribution of the Zeeman energy 
to spin-triplet excitation energies is ignored,
because it gives only constant shifts for excitation energies.

The HFB spectrum shows quantitative 
agreement with the ED results for large $\Delta_{\rm SAS}$ 
as $\Delta_{\rm SAS}/E_{C} = 0.7$.
On the other hand, for small $\Delta_{\rm SAS}$ 
the agreement between HFB and ED results becomes bad.
For example, for $\Delta_{\rm SAS}/E_{C} = 0.6$,
the spin-triplet HFB spectrum shows 
a mode softening in the long wavelength limit
overestimating the stability of the canted antiferromagnetic phase
(shown in the inset of Fig.~\ref{fig:spectrum}).
We note that similar results are obtained for 
ten-electron systems.


In our HFB theory 
a spin-unpolarized (SU) and pseudospin-polarized (PP) state 
$|\Psi_{\rm SU-PP}\rangle = \prod_{m}s_{m\uparrow}^{\dag}
s_{m\downarrow}^{\dag}|0\rangle$
is chosen 
as the reference state approximating the ground state.
This state is the vacuum state of 
$C_{KN}$, $D_{KN}$, $F_{KN}$, $H_{KN}$,
and in the HFB approximation 
these operators are treated as bosons.
They are transformed by a series of Bogoliubov transformations and
the ground state is obtained by applying these unitary transformations 
to $|\Psi_{\rm SU-PP}\rangle$.
Then the ground state is characterized as the vacuum state of
transformed bosons, $R_{KN}$, $S_{KN}$, $T_{KN}$ (three components of 
spin-triplet excitation), and $U_{KN}$ (spin-singlet one).
Thus our HFB theory can systematically describe not only 
spin-triplet and spin-singlet excitations but also 
the ground state wavefunction.
This is in striking contrast to the ambiguities in the HF theory.

\begin{figure}
\includegraphics[scale=.60]{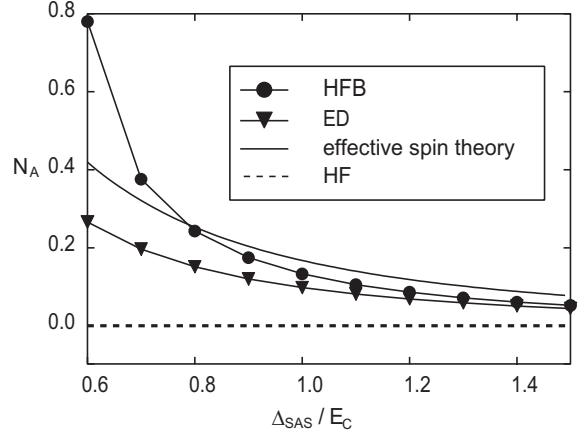}
\caption{The average number of electrons occupying antisymmetric 
single-particle states in the ground state 
is shown as a function of $\Delta_{\rm SAS}/E_{\rm C}$
for eight electrons and $d/l_{\rm B}=1$.
\label{fig:numanti}}
\end{figure}

In order to show clearly that the ground state properties are well 
described in our theory, the average number of 
electrons occupying antisymmetric single-particle states 
in ground state ($N_{A}$) is shown in Fig.\ref{fig:numanti}.
In the HFB theory, this quantity is given by
\[
N_{A}=\frac{1}{2}\sum_{K=0}^{2S}(2K+1)\left[3 \sinh ^2 
\biggl (\frac{\phi_{K}}{2}\biggl)
+\sinh ^2 \biggl (\frac{\varphi_{K}}{2} \biggl ) \right]. 
\label{eqn:nanti}
\]
In the figure, calculated results by the ED, 
the effective spin theory, and the HF method are also shown
in comparison with our result.
It is found that for large tunneling energies 
the HFB approximation does reproduce the ED 
result better than the effective spin theory and the HF does. 

On the other hand, for small tunneling energies as 
$\Delta_{\rm SAS}/E_{C} \leq 0.6$, 
the discrepancy between the ED and HFB theory becomes apparent
and the effective spin theory shows a better agreement 
with the ED result than the HFB theory.
This indicates that in small tunneling-energy region
another reference state describing pseudospin correlations better 
is needed in our HFB theory.

\section{Summary}
Using the Hartree-Fock-Bogoliubov approximation,
we have constructed an effective Hamiltonian for
bilayer $\nu=2$ QH systems.
This Hamiltonian preserves the spin rotational symmetry 
and gives both spin-singlet and spin-triplet excitations
systematically in contrast to the Hartree-Fock method.
In particular, 
in the large tunneling-energy region, 
our HFB theory describes the bilayer $\nu=2$ QH system better 
than other approximate theories. 
The ground-state properties are well described
by our theory, too.

T.N. and A.S. acknowledge support by Grant-in-Aid for Scientific Research
(Grant No.14740181 and No.14340088) by the Ministry of Education, 
Culture, Sports, Science and Technology of Japan, respectively.


\begin{thebibliography}{9}
\bibitem{QHE} {\it Quantum Hall Effect}, 
edited by R. E. Prange, S. M. Girvin 
(Springer-Verlag, New York, 1987).

\bibitem{QHE2} {\it Perspective in Quantum Hall Effects}, 
edited by S. Das Sarma, A. Pinczuk 
(Wiley, New York, 1997).

\bibitem{nu2HF1} S. Das Sarma, S. Sachdev, L. Zheng, 
Phys. Rev. Lett. {\bf 79}, 917 (1997)
; Phys. Rev. B {\bf 58}, 4672 (1998).

\bibitem{nu2HF2} A. H. MacDonald, R. Rajaraman, 
T. Jungwirth, Phys. Rev. B {\bf 60}, 8817 (1999).

\bibitem{est} E. Demler, S. Das Sarma, Phys. Rev. Lett. 
{\bf 82}, 3895 (1999).

\bibitem{nu2ED} J. Schliemann, A. H. MacDonald, 
Phys. Rev. Lett. {\bf 84}, 4437 (2000).

\bibitem{nu2exp1} A. Sawada {\it et al.}, 
Phys. Rev. Lett. {\bf 80}, 4534 (1998).

\bibitem{nu2exp2} V. Pellegrini {\it et al.}, 
Phys. Rev. Lett. {\bf 79}, 310 (1997); Science {\bf 281}, 799 (1998).

\bibitem{nu2exp3} V. S. Kharapai {\it et al.}, 
Phys. Rev. Lett. {\bf 84}, 725 (2000).

\bibitem{ring} On the Hartree-Fock-Bogoliubov theory,
see {\it The Nuclear Many-body Problem}, 
by P. Ring, P. Schuck (Springer-Verlag, Berlin, 1980).

\bibitem{nuclear} N. N. Bogoliubov, Uspekhi fiz. Nauk. 
{\bf 67}, 549 (1959) 
[translation: Soviet Phys.-Uspekhi {\bf 67}, 236 (1959)];
J. G. Valatin: Phys. Rev. {\bf 122}, 1012 (1961).

\bibitem{HFB} T. Nakajima, H. Aoki, Phys. Rev. B 
{\bf 56}, R15549 (1997).

\bibitem{sphere} Spherical systems are convenient to the theoretical 
extension to the fractional filling
based on the composite-fermion picture: 
X.G. Wu, J.K. Jain, Phys. Rev. B {\bf 49}, 7515 (1994);
T. Nakajima, H. Aoki, Phys. Rev. B {\bf 52}, 13780 (1995).


\end{thebibliography}
\end{document}